\documentclass[11pt, A4]{article}
\usepackage{graphicx}
\usepackage{epstopdf}
\usepackage{amsmath}
\usepackage{amssymb}
\usepackage{amsfonts}
\usepackage{amsthm}
\usepackage[usenames]{color}
\usepackage{array}
\usepackage{tabularx}
\usepackage{booktabs}           
\usepackage{caption}
\usepackage[usenames,dvipsnames,svgnames,table]{xcolor}
\usepackage[utf8]{inputenc}
\usepackage{lmodern}
\usepackage{verbatim}
\usepackage{amsmath}
\usepackage{amsfonts}
\usepackage{amssymb}
\usepackage{eso-pic}
\usepackage{graphicx}
\usepackage[final]{pdfpages}
\usepackage{pdfpages}
\usepackage[usenames,dvipsnames,svgnames,table]{xcolor}
\usepackage{mathtools}
\usepackage{empheq}

\usepackage[
      colorlinks=true,
      linkcolor=blue,
      urlcolor=blue,
      filecolor=blue,
      citecolor=red,
      pdfstartview=FitV,
      pdftitle={},
      pdfauthor={},
      pdfsubject={},
      pdfkeywords={},
      pdfpagemode=None,
      bookmarksopen=true
]{hyperref}

\usepackage{empheq}
\usepackage{mathtools}
\usepackage{epsfig}
\usepackage{cancel}
\usepackage{hyperref}

\def\beq{\begin{eqnarray}}
\def\eeq{\end{eqnarray}}

\def\e{\epsilon}

\def\be{\begin{equation}}
\def\ee{\end{equation}}
\def\bea{\begin{eqnarray}}
\def\eea{\end{eqnarray}}

\def\nn{\nonumber}

\definecolor{X}{rgb}{0,0,1}
\definecolor{Y}{rgb}{1,0,0}
\definecolor{Z}{rgb}{0,51,0}

\newcommand{\sbr}[1]{\left[#1\right]}

\newcommand{\br}[1]{\left(#1\right)}

\newcommand{\ab}{{\alpha\beta}}

\newcommand{\mn}{{\mu\nu}}
\newcommand{\rs}{{\rho\sigma}}

\newcommand{\n}{\nabla}

\renewcommand{\thefootnote}{\fnsymbol{footnote}}

\begin{document}

\begin{centering}

\thispagestyle{empty}


{\LARGE \textsc{A Note on Quantum fields in conformally flat space-times}} \\

 \vspace{0.8cm}

{\large 
 Swayamsidha Mishra$^{1,3}$, Sudipta Mukherji$^{2,3}$, and Yogesh K.~Srivastava$^{1,3}$}
\vspace{0.5cm}

\begin{minipage}{.9\textwidth}\small  \begin{center}
${}^{1}${National Institute of Science Education and Research (NISER), \\ Bhubaneswar, P.O. Jatni, Khurda, Odisha, India 752050}\\
  \vspace{0.5cm}
$^2$Institute of Physics, Sachivalaya Marg, \\ Bhubaneswar, Odisha, India 751005 \\
  \vspace{0.5cm}
$^3$Homi Bhabha National Institute, Training School Complex, \\ Anushakti Nagar, Mumbai India 400085 \\
  \vspace{0.5cm}
{\tt swayamsidha.mishra@niser.ac.in, mukherji@iopb.res.in, yogeshs@niser.ac.in}
\\ $ \, $ \\

\end{center}
\end{minipage}

\end{centering}

\renewcommand{\thefootnote}{\arabic{footnote}}

\begin{abstract}
We develop a technique relating scalar fields with different masses in different conformally 
flat space-times. We apply this technique to the case of FRW space-times, with $k=\pm 1,0$, and discuss several examples. We also study various 
energy 
conditions and discuss how they constrain the space-times related by this technique. 
We calculate the two point scalar correlator in the radiation dominated universe with
a hyperbolic spatial section from the one in the Milne universe using the above mapping. Finally, we consider trace anomaly and renormalized stress tensor for conformally flat space-times, especially Milne and radiation dominated universe (with $k=-1$)  using the transformation.
\end{abstract}

\newpage

\setcounter{equation}{0}

 Scalar field propagating in homogeneous and isotropic cosmological space-times has been extensively studied \cite{Birrell:1982ix}, 
\cite{parker}. It is useful in the study of cosmological perturbations, in studies of  particle production in cosmology and in quantum 
gravitational contexts. It has been known that Friedmann-Robertson-Walker (FRW) metrics are conformally flat as  evidenced by the vanishing 
of Weyl tensor for these 
space-times. In this short note, we develop a technique which utilizes conformal flatness to relate scalar fields propagating in two 
different FRWs. For flat FRW, a study was recently carried out in \cite{Lochan:2018pzs}. Our purpose here is to generalize this work in 
several ways.  

Let us consider two scalar fields $\phi$ and $\phi^\prime$ of mass parameters $m$ and $m^\prime$ respectively, propagating in $4$-dimensional 
conformally flat space-times endowed with metrics $g_{\mu\nu} = \Omega \eta_{\mu\nu}$ 
and 
$g^\prime_{\mu\nu} = \Omega^\prime \eta_{\mu\nu}$. To start with we assume that the scalars are minimally coupled to gravity. Then if
\begin{equation}
\Omega^\prime = F^2 \Omega,
\label{massrelation}
\end{equation}
for some space-time dependent function $F$ satisfying 
\begin{equation}
\Box_\Omega F - m^2 F + m'^2 F^3 = 0.
\label{massrelation1}
\end{equation}
then it follows that the  fields $\phi$ and $\phi^\prime$ are related 
by a field redefinition of the form
\begin{equation}
\phi^\prime = F^{-1} \phi.
\label{fieldrelation}
\end{equation} 
To check this, one starts with the Klein-Gordon equation for $\phi$
\begin{equation}
 \Box_{\Omega} \phi - m^2 \phi = 0.
\end{equation}
Noting that
\begin{equation}
\Box_{\Omega} \phi = \frac{1}{\sqrt{-g}} \partial_\mu({\sqrt{-g}} g^{\mu\nu} \partial_\nu \phi) = \frac{1}{\Omega^2} \partial_\mu\Big(\Omega^2 
\frac{\eta^{\mu\nu}}{\Omega} \partial_\nu \phi \Big),
\end{equation}
the KG equation can be rewritten as
\begin{equation}
\partial_\mu\Big(\frac{\Omega'}{F^2} \eta^{\mu\nu} (\phi^\prime\partial_\nu F + F \partial_\nu \phi')\Big) - \frac{\Omega'^2}{F^3} m^2 \phi' = 0.
\end{equation}
Here we have used (\ref{massrelation}) and (\ref{fieldrelation}).
It then immediately follows that 
\begin{equation}
 \Box^\prime_{\Omega'} \phi^\prime - {m^\prime}^2 \phi^\prime = 0,
\end{equation}
holds if (\ref{massrelation1}) is satisfied. If we work this out for the spatially flat FRW space-times with conformal scale factor $\Omega$, 
namely for the metric
\begin{equation} 
ds^2 = \Omega(\xi)(-d\xi^2 + dx^2 + dy^2 + dz^2),
\end{equation}
and the other metric with $\Omega$ replaced by $\Omega^\prime$, the equation relating masses (\ref{massrelation1})
reduces to 
\begin{equation}  
\frac{1}{F}(\partial_\xi\Omega) (\partial_\xi F) + \frac{\Omega}{F}\partial^2_\xi F + m^2\Omega^2 = {m^\prime}^2 \Omega^2 F^2.
\label{massrelation2}
\end{equation}
For the power law expansions, the consequence of the above correspondence is particularly enlightening and it 
leads to the following result. 
Consider a massless scalar in a space-time with $\Omega(\xi) = \xi^{-2q}$. It maps through (\ref{massrelation2}) to 
a massive scalar field of mass $m^{\prime 2}= (1 -q)(2+q)$ on a deSitter space with $\Omega^\prime(\xi) = \xi^{-2}$
where $F = \xi^{q-1}$. This result and it's further consequences thereof were recently analyzed in \cite{Lochan:2018pzs}.

The purpose of this note is to generalize the above correspondence in several ways. First, we extend it for the spatially non-flat 
FRW geometries in any dimensions. Subsequently, an  analogous mapping for the non-minimally coupled scalars is worked out. Various
energy conditions on the matter energy-momentum tensor (that sources the FRW space-time) put restrictions on the function $F$. 
We analyze these constraints. Finally, we end with some illustrative examples. One of which is the following. A massless minimally coupled 
scalar in a radiation dominated FRW geometry with a hyperbolic spatial section is related, by field redefinition, to a massless minimally 
coupled scalar in the Milne geometry. This, in turn, provides a connection between their quantum correlators which we subsequently 
analyze. Finally, we study the relationship between trace anomaly and
renormalized stress tensors for conformally coupled
scalars in conformally flat space-times, especially the relationship between the radiation dominated FRW and the Milne space-time.

\bigskip

\bigskip

\noindent {\bf Map for spatially non-flat FRW:} ~~We start with $k = -1$ open FRW metric in {\it $n$ dimensions}.
The metric has the form
\begin{equation}
ds^2= a^2(\eta)\br{-d\eta^2+d\chi^2+\sinh^2\chi d\Omega_{n-2}^2},
\label{HH}
\end{equation}
where $d\Omega_{n-2}^2$ is the metric on a unit $n-2$ sphere. For our purpose, it will be convenient to express
this metric in a conformally flat form \cite{Iihoshi:2007uz}
\begin{equation}
ds^2= a^2(\xi, r)A^2(\xi,r)\br{-d\xi^2+dr^2+r^2d\Omega_{n-2}^2},
\label{HH1}
\end{equation}
where the new coordinates are given by
\begin{eqnarray}
&&\xi = \frac{1}{\coth\frac{\eta+\chi}{2}+c}+\frac{1}{\coth\frac{\eta-\chi}{2}+c},\label{H3}\\
&&r = \frac{1}{\coth\frac{\eta+\chi}{2}+c}-\frac{1}{\coth\frac{\eta-\chi}{2}+c},\label{H4}
\end{eqnarray}
and 
\begin{equation}
A^2 = \frac{\br{\coth\frac{\eta+\chi}{2}+c}^2\br{\coth\frac{\eta-\chi}{2}+c}^2\sinh^2\chi}{\br{\coth\frac{\eta-\chi}{2}-\coth\frac{\eta+\chi}{2}}^2}.
\label{H33}
\end{equation}
Here $c$ is an arbitrary constant.
In the following, we will call $\Omega(\xi, r) = a^2(\xi, r)A^2(\xi, r)$. For the $k = 1$ closed FRW metric, we need to replace 
all the hyperbolic functions in (\ref{HH}) - (\ref{H33}) by the corresponding trigonometric functions.
In this way of writing, $k = 0$ Friedmann universe can also be accommodated simply by fixing  $A = 1$. $\Omega$ then becomes only a function 
of $\xi$.

To find an equation analogous to (\ref{massrelation2}), we take 
the following relations between $\phi$, $\phi^\prime$ and $\Omega$, $\Omega^\prime$.  
\bea
&& \phi^\prime = F^{-1}\phi,\\
&& \Omega' = F^{2p}\Omega.
\label{transf}
\eea
Here $p$ is some constant related to the dimension $n$ of the space, which we will evaluate while finding out the relation between $m$ and 
$m^\prime$.
Proceeding as before, Klein-Gordon equation $(\square_{\Omega} -m^2)\phi = 0$ can be written as   
\begin{eqnarray}
&&\partial_\mu\left(\Omega'^{(n-2)/2}\eta^{\mu\nu}\partial_\nu\phi^\prime\right) + \frac{2-(n-2)p}{F}\Omega'^{(n-2)/2}\eta^{\mu\nu}(\partial_\mu
 F)(\partial_\nu\phi^\prime)\nonumber\\
&&~~~~~~~+ F^{(n-2)p-1}\phi^\prime\partial_\mu\left(\frac{\Omega'^{(n-2)/2}}{F^{(n-2)p}}\eta^{\mu\nu}\partial_\nu F\right)
-\frac{\Omega'^{n/2}}{F^{2p}}m^2\phi^\prime = 0
\end{eqnarray}
The K-G equation for $\phi^\prime$ is
\begin{equation}
\partial_\mu(\Omega'^{(n-2)/2}\eta^{\mu\nu}\partial_\nu\phi^\prime)-\Omega'^{n/2}{m^\prime}^2\phi^\prime=0.
\end{equation}
Comparing the above two equations we reach at
\bea
{m^\prime}^2\phi^\prime = \frac{m^2}{F^{2p}}\phi^\prime -\frac{2-(n-2)p}{F^{2p+1}\Omega}\eta^{\mu\nu}(\partial_\mu
F)(\partial_\nu\phi^\prime)-\frac{F^{-2p-1}}{\Omega^{n/2}}\phi^\prime\partial_\mu\left(\Omega^{(n-2)/2}\eta^{\mu\nu}\partial_\nu F\right).
\eea
In order to get a relation between the parameters $m$ and $m^\prime$, we choose $p$ such that the term containing $\partial_\nu\phi^\prime$ in 
the above equation vanishes. This gives
\be
p= \frac{2}{(n-2)}.
\label{pvalue}
\ee
Putting this value of $p$, we finally obtain 
\be
{m^\prime}^2 = \frac{m^2}{F^{4/(n-2)}}-\frac{1}{F^{(n+2)/(n-2)}\Omega^{n/2}}\partial_\mu
\left(\Omega^{(n-2)/2}\eta^{\mu\nu}\partial_\nu F\right) = \frac{m^2}{F^{4/(n-2)}}-\frac{1}{F^{(n+2)/(n-2)}}\Box F.
\ee
This relation is valid for all three values of $k$, namely for $k = 0, \pm 1$. Only the explicit form of $\Omega$ changes.  
In passing, we note that for $n = 4, k = -1$ the above relation reduces to
\begin{equation}
\frac{\partial^2_\xi F}{F}+\frac{\partial_\xi
\Omega\partial_\xi F}{F\Omega}-\frac{2\partial_r F}{Fr}-\frac{\partial^2_r F}{F}-
\frac{\partial_r\Omega\partial_r F}{F\Omega}+m^2\Omega={m^\prime}^2\Omega F^2.
\label{k-1mass}
\end{equation}
We will require this equation later. 

It is easy to see that we get same results if we start with the action 
instead of the equation of motion. For simplicity, we discuss only $4$-dimensional minimally coupled scalar case here. Consider the action for a  scalar field in a conformally flat space-time.
\bea
S &=& \int d^4x \sqrt{-g}\sbr{ - g^ {\mu\nu}\partial_\mu\phi\partial_\nu\phi- m^2\phi^2}\nn\\  
&=& \int d^4x \,\,\Omega^2(\xi,r)\sqrt{-\eta}\sbr{ - \Omega^{-1}\eta^{\mu\nu}\partial_\mu\phi\partial_\nu\phi- m^2\phi^2}. \nn
\eea
After the conformal transformation we get
\bea
S'  = \int d^4x \,\,\Omega'^2(\xi,r) \sqrt{-\eta}\sbr{ - \Omega'^{-1}\eta^{\mu\nu}\partial_\mu\phi^\prime\partial_\nu\phi^\prime- m^{\prime 
2}\phi^{\prime 2}}.\nn
\eea
We have the relation between the two masses (\ref{k-1mass}) along with the relation $\phi^\prime = F^{-1}\phi$ and
we can check the invariance of the action under the conformal transformation $\Omega' = F^2\Omega $.
It immediately follows that   
\be
S' = S + \int d^4x\,\partial_\mu\sbr{\sqrt{-\eta}\frac{\phi^2\Omega}{F}\partial^\mu F}. \nonumber
\ee

\bigskip

\bigskip

\noindent{\bf Non-minimally coupled scalars:} ~~The correspondence analogous to the above can be generalized for non-minimally coupled 
scalars as 
well.
Consider the Klein-Gordon equations
\begin{equation}
(\square_{\Omega} -m^2 -\varepsilon R)\phi  = 0,~~{\rm and}~~(\square^\prime_{\Omega'} -{m^\prime}^2 -\varepsilon R')\phi^\prime = 0,
\label{verep}
\end{equation}
where, $\epsilon$ is a constant.
Using (\ref{transf}) and (\ref{pvalue}), we find that the Ricci tensors are related by
\begin{eqnarray}
&&R'_{\alpha\beta}
= R_{\alpha\beta}-(n-2)\nabla_\alpha\nabla_\beta \ln F^{\frac{2}{n-2}}-g_{\alpha\beta}g^{\gamma\delta}\nabla_\gamma\nabla_\delta \ln 
F^{\frac{2}{n-2}}\nn\\
&&~~~~~~~~~~+(n-2)(\nabla_\alpha\ln
F^{\frac{2}{n-2}})(\nabla_\beta\ln F^{\frac{2}{n-2}})\nn\\
&&~~~~~~~~~~-(n-2)g_{\alpha\beta}g^{\gamma\delta}(\nabla_\gamma \ln F^{\frac{2}{n-2}})(\nabla_\delta \ln F^{\frac{2}{n-2}}).
\label{newRT}
\end{eqnarray}
Ricci scalars $R$ and $R^\prime$ are related 
as
\begin{eqnarray}
&&R'=  F^{-\frac{4}{n-2}} \Big[R- 2(n-1)g^{\alpha\beta}\nabla_\alpha\nabla_\beta \ln F^{\frac{2}{n-2}}\nn\\
 &&~~~~~~~~~~-(n-2)(n-1)g^{\alpha\beta}(\nabla_\alpha \ln F^{\frac{2}{n-2}})
(\nabla_\beta\ln F^{\frac{2}{n-2}})\Big].
\label{newRS}
\end{eqnarray}
Using this and performing a similar computation as before we find the masses are related by
\begin{equation}
{m^\prime}^2 = \frac{m^2 }{F^{\frac{4}{n-2}}} -
\left[1-\frac{4\varepsilon(n-1)}{(n-2)}\right]\frac{1}{F^{\frac{n+2}{n-2}}\Omega^{\frac{n}{2}}}\partial_\alpha\left(\Omega^{\frac{n-2}{2}}
\eta^{\alpha\beta}\partial_\beta  F\right).
\end{equation}
In the second equation of (\ref{verep}), we could have used different
constant $\varepsilon^\prime$. The extension is straightforward.

\bigskip

\bigskip

\noindent{\bf Energy Conditions:} ~~In order that the two metrics, one with conformal factor $\Omega$ and the other with $\Omega^\prime$, be 
the solutions 
of the Einstein equations, they will have to be supported by appropriate energy-momentum tensors $T_{\alpha\beta}$. 
We now work out the relations between $T_{\alpha\beta}$ and ${T^\prime}_{\alpha\beta}$ and 
further comment on the restrictions on the function $F$ coming from various energy conditions.
The Einstein equations are
\begin{equation}
T_{\alpha\beta} = R_{\alpha\beta} -\frac{1}{2}g_{\alpha\beta}R,~~T'_{\alpha\beta} = R'_{\alpha\beta} -\frac{1}{2}g'_{\alpha\beta}R'.\nn
\end{equation}
Now using (\ref{newRT}) and (\ref{newRS}), we find, 
\bea
 T'_{\alpha\beta } &&=  T_{\alpha\beta}-2\left[-\frac{1}{F^2}(\nabla_\alpha F)(\nabla_\beta F) +\frac{1}{F}\nabla_\alpha\nabla_\beta
F\right]+2g_{\alpha\beta}g^{\gamma\delta}\left[-\frac{1}{F^2}(\nabla_\gamma F)(\nabla_\delta F) +\frac{1}{F}\nabla_\gamma\nabla_\delta F\right]\nn\\
 &&~~~~~~~ +\frac{4}{(n-2)F^2}(\nabla_\alpha F)(\nabla_\beta F)+\frac{2(n-3)}{(n-2)F^2}g_{\alpha\beta} g^{\gamma\delta}(\nabla_\gamma F) 
(\nabla_\delta 
F)\nn\\
 &&=  T_{\alpha\beta} + \frac{2n}{(n-2)F^2}(\nabla_\alpha F)(\nabla_\beta F)-\frac{2}{F}\nabla_\alpha\nabla_\beta F\nn\\
&&~~~~~~~- \frac{2}{(n-2)F^2}g_{\alpha\beta}
g^{\gamma\delta}(\nabla_\gamma F) (\nabla_\delta F)+\frac{2}{F}g_{\alpha\beta} g^{\gamma\delta}\nabla_\gamma\nabla_\delta F.
\label{newT}
\eea

Two energy-momentum tensors which are related as above still need to satisfy the energy conditions to be
physically relevant. Simplest of the energy conditions is the null energy condition.
The null energy condition stipulates that for every future-pointing null vector field $k^\mu$
\be
\rho = T_{\mu\nu}k^\mu k^\nu \geq 0.
\ee
We now find out the constraints on the function $F$ for both $T^{\mu\nu}$ and $T'^{\mu\nu}$ to satisfy the above energy conditions. 
Note that if $k^\mu $ is a
null vector with respect to the unprimed metric then it is a null vector with respect to the primed metric as well. 
Using expressions relating old and new energy-momentum
tensors as in (\ref{newT}), we can write the constraint of null energy condition
for $4$-dimensional case as
\begin{equation}
T^\prime_{\alpha\beta}k^\alpha k^\beta = T_{\alpha\beta}k^\alpha k^\beta +\frac{4}{F^2}k^\alpha k^\beta(\nabla_\alpha F)(\nabla_\beta F) - 
\frac{2}{F}k^\alpha 
k^\beta\nabla_\alpha\nabla_\beta F \geq 0\nn
\end{equation}
In particular, if we start with $T_{\mu\nu}k^\mu k^\nu = 0$,
\begin{equation}
2k^\alpha k^\beta(\nabla_\alpha F)(\nabla_\beta F) - Fk^\alpha
k^\beta\nabla_\alpha\nabla_\beta F \geq 0.\nn
\end{equation}

The weak energy condition, on the other hand, tells us that for every future-directed timelike vector field $X^\mu$,
the matter density observed by the corresponding observer is always non-negative
\begin{equation}
\rho = T_{\mu\nu} X^\mu X^\nu \geq 0.\nn
\end{equation}
In our context, this gives
\bea
T'_{\alpha\beta} X^\alpha X^\beta &=& T_{\alpha\beta} X^\alpha X^\beta +\frac{4}{F^2}X^\alpha X^\beta (\nabla_\alpha F)(\nabla_\beta 
F) - \frac{2}{F}X^\alpha X^\beta\nabla_\alpha\nabla_\beta F \nn\\ 
&-& \frac{1}{F^2}
X^\alpha X^\beta g_{\alpha\beta}g^{\gamma\delta}(\nabla_\gamma F)(\nabla_\delta F)+ \frac{2}{F} X^\alpha X^\beta
g_{\alpha\beta}g^{\gamma\delta}\nabla_\gamma\nabla_\delta F \geq 0\nn
\eea 
Having reached thus far, in the rest of the note, we consider a few illustrative examples. In the following, for simplicity,  we will
restrict ourselves to the massless scalars.

\bigskip

\bigskip

\noindent{\bf Illustrative examples:} 
We show that a simple consequence of our earlier discussion is that a field of a massless scalar in a four dimensional Milne universe is related by field redefinition to that of one in a radiation dominated universe with hyperbolic spatial section in the same space-time dimensions.

We start with a Milne universe in four dimensions.
It has the form of (\ref{HH}), with $a(\eta) = e^{\eta}$ and $n = 4$. 
\begin{equation}
ds^2= e^{2\eta}\br{-d\eta^2+d\chi^2+\sinh^2\chi d\Omega_{2}^2}.\nn
\end{equation}
Via coordinate transformations
\begin{equation}
T = e^{\eta} ~{\rm cosh}\chi,~~R = e^{\eta}~{\rm sinh}\chi,
\end{equation}
it can be mapped to a flat metric. However, as can be readily seen from the above transformations,
Milne coordinates cover only the future directed light cone of the flat space-time.
In $\xi, r$ coordinates of (\ref{HH1}) - (\ref{H4}),
the line element can be written as
\bea
ds_M^2 &=& \frac{1}{\sbr{(1-\xi)^2-r^2}^2}\br{-d\xi^2+dr^2+r^2d\Omega_2^2}.
\eea
Therefore
\be
\Omega = \frac{1}{\sbr{(1-\xi)^2-r^2}^2}.
\ee
We have chosen $c = 1$ in (\ref{H3}) and (\ref{H4}).

We start with a massless scalar $\phi$ in this background and, as before, rewrite the K-G equation 
again for a massless field $\phi^\prime = F^{-1} \phi$ and $\Omega^\prime = F^2 \Omega$. We find that there
exists a solution of $F$ in (\ref{k-1mass}) for which $\Omega^\prime$ or in other words, $g_{\mu\nu}^\prime$
is that of a hyperbolic radiation dominated universe. It is easy to see that in this particular case, 
(\ref{k-1mass}) reads as 
\bea
 -\sbr{(1-\xi)^2-r^2}\partial_{\xi}^2 F &-& 4(1-\xi)\partial_\xi F +
\sbr{(1-\xi)^2-r^2}\partial_{r}^2F \nn\\
&+& 4r\partial_r F +\frac{2}{r}\sbr{(1-\xi)^2-r^2}
\partial_r F= 0.\nn
\eea
The general solution of the above equation is 
\begin{equation}
F = c_1 - c_2 \Big[(1 - \xi)^2 - r^2\Big],
\end{equation}
where $c_1$ and $c_2$ are two constants. For the special case of $c_1, c_2 = 1/2$, we get
\begin{equation}
\Omega^\prime = F^2 \Omega = \Big[ \frac{r^2 - \xi^2 + 2 \xi}{2 (1-\xi)^2 - 2 r^2}\Big]^2.
\end{equation}
Writing in terms of the hyperbolic coordinates as in (\ref{HH}), we get 
\begin{equation}
{ds^\prime}^2 = {\rm sinh}^2 \eta \Big(-d\eta^2 + d\chi^2 + {\rm sinh}^2\chi d\Omega_2^2\Big).
\end{equation}
This geometry solves the Einstein equation in the presence of a perfect fluid energy-momentum tensor with
the equation of state of a radiation dominated universe, namely 
$p = \omega \rho$ with $ \omega = 1/3$. We therefore conclude that a massless 
scalar field 
in a radiation dominated open FRW universe is related to another massless scalar in a Milne space-time
via a field redefinition. In particular, we find
\begin{equation}
\phi' (\eta) = \frac{e^{\eta}}{{\rm sinh}\eta} ~\phi(\eta).
\end{equation}
This correspondence is useful because it allows us to relate correlators in two different
geometries. Consider, for example, a massive scalar of mass $m$ in Milne space-time. The normalized solution is given by
\cite{Yamamoto:1994te}
\begin{equation}
\hat \phi(\eta, \chi, \Omega) = \int_0^\infty dp \sum_{lm} \Big[ \frac{\sqrt{\pi}}{2} \e^{\pi p/2} e^{-\eta}
H_{ip}^{(2)} (me^{\eta}) Y_{plm}(\chi, \Omega) \hat b_{plm} + {\rm h.c.}\Big],\nn
\end{equation}
where $H_{ip}^{(2)} (me^{\eta})$ is the Hankel function and $Y_{plm}$ is given by
\begin{eqnarray}
&&Y_{plm}(\chi, \Omega) = f_{pl}(\chi) Y_{lm}(\Omega),\nn\\
&&f_{pl} = \frac{\Gamma(ip + l+1)}{\Gamma(ip)} \frac{1}{\sqrt{{\rm sinh}\chi}} P^{-l - 1/2}_{ip - 1/2} ({\rm cosh\chi}).\nn
\end{eqnarray}
In the above equations, $Y_{lm}, \Gamma$ and $ P^{-l - 1/2}_{ip - 1/2}$ are the spherical harmonics on the unit sphere, the 
gamma function and the 
associated Legendre function respectively. Here we have taken the solution which corresponds to the positive 
frequency function
with respect to the Minkowski vacuum. The $Y_{plm}$ satisfy the completeness condition \cite{Sasaki:1994yt}
\begin{equation} 
\sum_{lm} Y_{plm}(\chi, \Omega) Y_{plm}(\chi^\prime, \Omega^\prime) = \frac{p~{\rm sin} p\zeta}{2\pi^2 {\rm sinh}\zeta},\nn
\end{equation}
$\zeta$ defined in (\ref{zetadef}).
The propagator is then
\begin{eqnarray}
&&\langle [\hat\phi(\eta, \chi, \Omega) \hat\phi(\eta^\prime, \chi^\prime, \Omega^\prime)]\rangle_{\rm Milne} \nn\\
&&~~~~~= \frac{\pi}{4 e^{\eta + \eta^\prime}} \int_{0}^{\infty} dp e^{\pi p} H_{ip}^{(2)} (me^\eta) H_{ip}^{(2)*} (m e^{\eta^\prime}) 
\sum_{lm} 
Y_{plm}(\chi, \Omega)
Y_{plm}^{*}(\chi^\prime, \Omega^\prime) \nn\\
&&~~~~~= \frac{1}{8 \pi e^{\eta+\eta^\prime} {\rm sinh}\zeta } \int_{0}^{\infty} dp ~e^{\pi p} p~{\rm sin}p\zeta
~H_{ip}^{(2)} (m e^\eta) H_{ip}^{(2)*} (m e^{\eta\prime})\nn\\
&&~~~~~= \frac{1}{8 \pi e^{\eta +\eta^\prime} {\rm sinh}\zeta } \int_{0}^{\infty}dp  ~p~{\rm sin}p\zeta ~H_{ip}^{(2)} (me^\eta) H_{ip}^{(1)} (m 
e^{\eta^\prime})\nn\\
&&~~~~~= - \frac{1}{16 \pi e^{\eta+\eta^\prime} ~{\rm sinh}\zeta} \frac{\partial}{\partial \zeta}\Bigg[ \int_{-\infty}^{\infty} dp e^{-ip\zeta}
H_{ip}^{(2)}(m e^\eta) H_{ip}^{(1)}(m e^{\eta^\prime})\Bigg].
\label{hankel2}
\end{eqnarray}
In the above equations,
\begin{eqnarray}
&&{\rm cosh}\zeta = {\rm cosh}\chi ~{\rm cosh}\chi^\prime - {\rm sinh}\chi ~{\rm
sinh}\chi^\prime {\rm cos}\omega,\nn\\
&&{\rm cos}\omega = {\rm cos}\theta ~{\rm cos}\theta^\prime + {\rm sin}\theta~{\rm sin}\theta^\prime~{\rm cos}(\phi - \phi^\prime).
\label{zetadef}
\end{eqnarray}
Now using the identity\footnote{See equation (II.9) of \cite{diSessa:1974ve}.}
\begin{equation}
\int_{-\infty}^{\infty} dp e^{-ip\zeta}
H_{ip}^{(2)}(m e^\eta) H_{ip}^{(1)}(m e^{\eta^\prime}) = - 2 i H_{0}^{(2)}\Big( {\sqrt{m^2 e^{2\eta} + m^2 e^{2\eta^\prime} - 2 
m^2e^{\eta+\eta^\prime}
~{\rm cosh}\zeta - i(\eta - \eta^\prime)\epsilon}}\Big)\nn,
\end{equation}
in (\ref{hankel2}), we arrive at the Feynman propagator
\begin{eqnarray}
\langle T[\hat\phi(\eta, \chi, \Omega) \hat\phi(\eta^\prime, \chi^\prime, \Omega^\prime)]\rangle_{\rm Milne}
= \Big(\frac{i m}{8\pi {\sqrt{-\sigma^2}} }\Big)  H_{1}^{(2)}\Big( m{\sqrt{-\sigma^2}}\Big)\nn.
\end{eqnarray}
where
\begin{equation}
\sigma^2 =  -e^{2\eta} - e^{2\eta^\prime} + 2 e^{\eta+\eta^\prime}
~{\rm cosh}\zeta + i(\eta - \eta^\prime)\epsilon.\nn
\end{equation}
In the limit $m \rightarrow 0$, the correlator reduces to
\begin{equation}
\langle T[\hat\phi(\eta, \chi, \Omega) \hat\phi(\eta^\prime, \chi^\prime, \Omega^\prime)]\rangle_{\rm Milne} =
\frac{1}{4\pi^2 \sigma^2}.
\nn
\end{equation}
Note that it has the expected structure of the flat space-time propagator restricted to the Milne region. 
The correlator for the  radiation dominated universe in the Milne vacuum is then\footnote{It is argued 
in \cite{Redmount:1999ps} that the Milne vacuum is the natural choice of vacuum for a radiation dominated Friedmann universe.} 
\begin{equation}
\langle T[\hat\phi^\prime(\eta, \chi, \Omega) \hat\phi^\prime(\eta^\prime, \chi^\prime, \Omega^\prime)]\rangle_{\rm Radiation}
=  \frac{e^{(\eta + \eta^\prime)}}{4\pi^2 \sigma^2~{\rm sinh}\eta~{\rm sinh}\eta^\prime}.
\nn
\end{equation}
Note that both the radiation dominated universe and the Milne universe satisfy the energy conditioned mentioned previously.

Many such mappings can be constructed simply by solving the equation for $F$. 
Consider a FRW with conformal scale factor $a(\eta)$ and new one with  a scale factor $b(\eta)= F(\eta) a(\eta)$, then the equation 
relating masses can be cast to  
\be
m^{\prime^2} = \frac{m^2}{F^2}+ \frac{2}{F^3a^3} \partial_\eta a 
\partial_\eta F +\frac{1}{F^3a^2}\partial^2_\eta F.
\label{nokchange}
\ee
When the scalar fields in $a(\eta)$ FRW and $b(\eta)= F (\eta)a(\eta)$ FRW are both massless,
general solution for the above equation is $ F(\eta) = c_1 + c_2\int^\eta \frac{d\eta'}{a^2(\eta')}$.
Consider the  $k=0$ case with $a(\eta)= \eta^\alpha$, $\alpha = \frac{2}{3\omega + 1}$. Here $\omega$ is the parameter 
appearing in the linear equation of state, $p=\omega\rho$ as before. In this case, equation (\ref{nokchange}) leads to
\bea
b(\eta) = F(\eta)a(\eta) =  c_2 \eta^{\frac{(3\omega - 1)}{(3\omega + 1)}}.  \nn
\eea
We see from this that starting with a FRW with the  dust ($\omega = 0$), we get one with a cosmological constant ($\omega = -1$) and 
vice-versa. Similarly, starting with a radiation dominated FRW ($\omega = 1/3$), we get  the flat Minkowski space and vice-versa.

\bigskip

\bigskip

\noindent{\bf Quantum Stress Tensor and Trace anomaly:} ~~ In the light of our previous analysis,
in this section, we briefly consider the stress tensor and 
the trace anomaly for conformally flat space-times. For simplicity, we consider conformally coupled scalar with $\epsilon =\frac{1}{6}$. 
For this case, the system is conformally trivial and we verify that the trace anomaly 
and the stress tensors are the expected ones. In particular, 
we verify the relation between the stress tensor of Milne and the radiation dominated universe\footnote{Since for both the geometries Ricci scalars vanish, the choice 
of $\epsilon$ is immaterial.}. 

Formula for the trace anomaly for conformally coupled massless scalar field in  a general FRW background, after converting to mostly plus signature,  
is\cite{Bunch:1977sq}
\bea
\langle T^\mu_{\,\,\,\mu}\rangle = (2880\pi^2)^{-1}\sbr{\Box R +\br{R^\mn R_\mn -\frac{1}{3}R^2}}.
\eea
For the radiation dominated universe, it evaluates to 
\bea
\langle T^\mu_{\,\,\mu}\rangle = 
\frac{{\rm cosech}^8\eta}{240\pi^2}.
\label{anom}
\eea

Next we derive the above by using the transformation of the trace-anomaly equation under a conformal transformation. 
Since the trace-anomaly equation is a local equation which is  independent of the state, we do not need to worry about the transformation of states.
The vev of the trace of the renormalized stress energy tensor transforms under a conformal transformation of the metric so that we can relate them by the conformal 
factor. After a lengthy calculation, we get 
\begin{align}
\Rightarrow \langle T_\mu^{\prime\,\,\mu}\rangle &= \frac{1}{F^4}\langle T_\mu^{\,\,\mu}\rangle  +(2880\pi^2)^{-1}\Big[ -\frac{2}{F^5}g^\mn (\n_\mu F)(\n_\nu R)     -\frac{4}{F^5}R^\mn\n_\mu\n_\nu F    +\frac{8}{F^6}R^\mn (\n_\mu F)(\n_\nu F)  \nn\\
&   +\frac{4}{F^6}(\n^\mu\n^\nu F)(\n_\mu\n_\nu F)    -\frac{16}{F^7}(\n^\mu\n^\nu F)(\n_\mu F)(\n_\nu F)     +\frac{12}{F^8}[(\n^\mu F)(\n_\mu F)]^2\nn\\
& -\frac{32}{F^7}g^\mn (\n_\mu F)(\n_\nu F)(\Box F)     +\frac{24}{F^6}g^\mn(\n_\mu F)(\n_\nu\Box F)        +\frac{14}{F^6}(\Box F)^2      -\frac{6}{F^5}\Box^2 F \Big]
\end{align}
For Milne to radiation dominated case, all the components of the Ricci tensor and the Ricci scalar vanish (for Milne), hence $\langle T_\mu^{\,\,\mu}\rangle = 0 $. 
Further using
$\Box F = 0$,
we get for the radiation dominated FRW,
\bea
\langle  T_\mu^{\prime\,\,\mu}\rangle 
=   (2880\pi^2)^{-1}\frac{4}{F^8}\sbr{ F^2(\n^\mu\n^\nu F)(\n_\mu\n_\nu F)    -4F(\n^\mu\n^\nu F)(\n_\mu F)(\n_\nu F)    +3\sbr{(\n^\mu F)(\n_\mu F)}^2  }.\nn
\eea
Finally, using $F= e^{-\eta} \sinh\eta$, we get
\bea
\langle T^{\prime\mu}_{\,\,\,\,\mu}\rangle&=& \frac{{\rm cosech}^8\eta}{240\pi^2},
\label{anom1}
\eea
which matches with the previous result (\ref{anom}). Now we consider the stress  tensor itself rather than just the trace of it. The renormalized vev of the stress 
energy tensor for conformally coupled scalar fields in conformally related space-times is given by Candelas and Dowker in 
\cite{Candelas:1978gf}. As discussed in that paper, the conformal vacua can be classified into two families, namely, the Rindler and 
the Minkowski families. Since both the Milne and the 
hyperbolic radiation 
dominated  universe belong 
to the same family, we can relate  $\langle T_\mu^{\,\,\nu}\rangle $ for these two metrics. The relation goes as 
\begin{align}
\langle T_\mu^{\prime\,\,\nu}\rangle &= \frac{1}{\sqrt{-g^\prime}}\sqrt{-g}\langle T_\mu^{\,\,\nu}\rangle_{NG} +   \frac{1}{2880\pi^2}\Bigg[a(s) 
\br{2\n^\prime_\mu\n^{\prime\nu} R^\prime -2R^\prime R_\mu^{\prime\,\,\nu} +\frac{1}{2}\delta_\mu^\nu R^{\prime 2} -
2\delta_\mu^\nu \Box^\prime R^\prime}\label{C-D-EM}\nn\\
& \qquad\qquad\qquad\qquad\qquad\qquad\qquad\quad + b(s)\br{\frac{2}{3}R^\prime R_\mu^{\prime\,\,\nu} -
R_{\mu\alpha}^\prime R^{\prime\nu\alpha} -\frac{1}{4}\delta_\mu^\nu R^{\prime 2} +\frac{1}{2}\delta_\mu^\nu R_\ab^\prime R^{\prime\ab}}\Bigg],\nn
\end{align}
where $\langle T_\mu^{\,\,\nu}\rangle_{NG}$ is only the non-geometrical part of $\langle T_\mu^{\,\,\nu}\rangle $ for the starting space-time 
(non-geometrical part arises because we are in the Rindler family instead of the Minkowski family). 
Constants $a(s)$ and $b(s)$ are -1/6 and 1 respectively for spin zero scalar fields. We can write the above formula in terms of $F$ as
\begin{align}
\langle T_\mu^{\prime\,\,\nu}\rangle =& \frac{1}{F^4} \sbr{\langle T_\mu^{\,\,\nu}\rangle_{NG}   +(2880\pi^2)^{-1}\br{ -\frac{1}{3}\n_\mu\n^\nu R  +\frac{1}{3}\delta_\mu^\nu\Box R   +  R R_{\mu}^{\,\,\nu}    -\frac{1}{3}\delta_\mu^\nu R^{2}        -R_{\mu\alpha}R^{\nu\alpha}     +\frac{1}{2}\delta_\mu^\nu R_{\ab} R^{\ab} }}\nn\\
&  +   (2880\pi^2)^{-1}\Big[   \frac{1}{F^5}(\n_\mu F)(\n^\nu R)     +\frac{1}{F^5}(\n^\nu F)(\n_\mu R)       -\frac{1}{F^5}\delta_\mu^\nu g^\rs(\n_\sigma F)(\n_\rho R) \nn\\
&  +\frac{2}{3F^6}R(\n_\mu F)(\n^\nu F)        -\frac{2}{3F^6}\delta_\mu^\nu g^\rs R(\n_\sigma F)(\n_\rho F)     +\frac{2}{F^6}R_\mu^{\,\,\nu}  g^\rs(\n_\rho F)(\n_\sigma F) \nn\\
& + \frac{2}{F^5}R_{\mu\alpha}\n^\nu\n^\alpha F       +\frac{2}{F^5}R^{\nu\alpha}(\n_\mu\n_\alpha F)     -\frac{2}{F^5}\delta_\mu^\nu R^{\ab}(\n_\alpha\n_\beta F) -\frac{4}{F^6}R_{\mu\alpha}(\n^\nu F)(\n^\alpha F)\nn\\
& - \frac{4}{F^6}R^{\nu\alpha}(\n_\mu F)(\n_\alpha F)      + \frac{4}{F^6}\delta_\mu^\nu R^{\ab}(\n_\alpha F)(\n_\beta F)    -\frac{4}{F^6}(\n_\mu\n_\alpha F)(\n^\nu\n^\alpha F)     \nn\\
&  +\frac{2}{F^6}\delta_\mu^\nu (\n_\alpha\n_\beta F)(\n^\alpha\n^\beta F)       +\frac{8}{F^7}(\n_\mu\n_\alpha F)(\n^\nu F)(\n^\alpha F)    +\frac{8}{F^7}(\n_\mu F)(\n_\alpha F)(\n^\nu\n^\alpha F)\nn\\
&   -\frac{8}{F^7}\delta_\mu^\nu (\n_\alpha\n_\beta F)(\n^\alpha F)(\n^\beta F)      -\frac{8}{F^8}(\n_\mu F)(\n_\alpha F)(\n^\nu F)(\n^\alpha F)    \nn\\
&  +\frac{5}{F^8}\delta_\mu^\nu (\n_\alpha F)(\n_\beta F)(\n^\alpha F)(\n^\beta F)      -\frac{4}{F^7}(\n_\mu\n^\nu F)g^\rs(\n_\rho F)(\n_\sigma F)    \nn\\
& +\frac{20}{F^7}(\n_\mu F)(\n^\nu F)\Box F    -\frac{12}{F^7}\delta_\mu^\nu g^\rs (\n_\rho F)(\n_\sigma F)\Box F    -\frac{8}{F^6}(\n_\mu F)(\n^\nu\Box F)     \nn\\
&-\frac{8}{F^6}(\n^\nu F)(\n_\mu\Box F)        +\frac{10}{F^6}\delta_\mu^\nu g^\rs (\n_\sigma F)(\n_\rho\Box F)     -\frac{4}{3F^5}R\n_\mu\n^\nu F    +\frac{4}{3F^5}\delta_\mu^\nu R\Box F\nn\\
& -\frac{4}{F^5} R_{\mu}^{\,\,\nu}\Box F     +\frac{2}{F^6}(\n_\mu\n^\nu F)\Box F    +\frac{3}{F^6}\delta_\mu^\nu(\Box F)^2    +\frac{2}{F^5}\n_\mu\n^\nu\Box F     -\frac{2}{F^5}\delta_\mu^\nu\Box^2 F \Big]
\end{align}

As mentioned previously, since the Milne and the radiation dominated space-times belong to the same family, we can apply the above formula 
to relate the corresponding tensors 
\footnote{$\langle T_\mu^{\,\,\nu}\rangle_{NG} = \langle T_\mu^{\,\,\nu}\rangle$ for Milne with vanishing trace.}. Noting that $\Box F=0$, we reach at
\bea
\langle T^{\prime\,\,\nu}_{\mu}\rangle &=&  \frac{1}{F^4}\langle T_\mu^{\,\,\nu}\rangle    +(2880\pi^2)^{-1}\Big[ -\frac{4}{F^6}(\n_\mu\n_\alpha F)(\n^\nu\n^\alpha F)    +\frac{2}{F^6}\delta_\mu^\nu (\n_\alpha\n_\beta F)(\n^\alpha\n^\beta F)  \nn\\
&& 
+\frac{8}{F^7}(\n_\mu\n_\alpha F)(\n^\nu F)(\n^\alpha F)  +\frac{8}{F^7}(\n_\mu F)(\n_\alpha F)(\n^\nu\n^\alpha F)  \nn\\
&& -\frac{8}{F^7}\delta_\mu^\nu (\n_\alpha\n_\beta F)(\n^\alpha F)(\n^\beta F) -\frac{4}{F^7}(\n_\mu\n^\nu F)g^\rs(\n_\rho F)(\n_\sigma F)  \nn\\
&&    -\frac{8}{F^8}(\n_\mu F)(\n_\alpha F)(\n^\nu F)(\n^\alpha F)    +\frac{5}{F^8}\delta_\mu^\nu (\n_\alpha F)(\n_\beta F)(\n^\alpha F)(\n^\beta F)        \Big]
\eea  
We see from the above equation that only the diagonal terms of $\langle T^{\prime\,\,\nu}_{\mu}\rangle $ are non-zero. Explicitly, these are 
\bea
\langle T^{\prime\,\,\eta}_{\eta}\rangle &=& \frac{1}{F^4}\langle T^{\,\,\eta}_{\eta}\rangle   - (2880\pi^2)^{-1} 3{\rm cosech}^8\eta   = 
(2880\pi^2)^{-1}\br{6{\rm cosech}^4\eta- 
3{\rm cosech}^8\eta}, \nn\\
\langle T^{\prime\,\,\chi}_{\chi}\rangle &=&\frac{1}{F^4}\langle T^{\,\,\chi}_{\chi}\rangle   + (2880\pi^2)^{-1} 5{\rm cosech}^8\eta   =(2880\pi^2)^{-1} 
\br{-2{\rm cosech}^4\eta+5{\rm cosech}^8\eta}, \nn\\
\langle T^{\prime\,\,\theta}_{\theta}\rangle &=&\frac{1}{F^4} \langle T^{\,\,\theta}_{\theta}\rangle    + (2880\pi^2)^{-1} 5{\rm cosech}^8\eta =(2880\pi^2)^{-1} 
\br{-2{\rm cosech}^4\eta+5{\rm cosech}^8\eta},\nn \\
\langle T^{\prime\,\,\phi}_{\phi}\rangle &=& \frac{1}{F^4} \langle T^{\,\,\phi}_{\phi}\rangle    + (2880\pi^2)^{-1} 5{\rm cosech}^8\eta =(2880\pi^2)^{-1} 
\br{-2{\rm cosech}^4\eta+5{\rm cosech}^8\eta}.\nn 
\eea
Here we have used the non-geometrical part \cite{Candelas:1978gf} for the Milne (which is equal to the total $\langle T_\mu^{\,\,\nu}\rangle$) given by
\be
\langle T_\mu^{\,\,\nu}\rangle_{NG} =\langle T_\mu^{\,\,\nu}\rangle   = (2880\pi^2)^{-1}\text{diag}\br{6e^{-4\eta},-2e^{-4\eta},-2e^{-4\eta},-2e^{-4\eta}}.
\ee
We can again calculate the trace
\bea 
\langle T^{\prime\,\,\mu}_{\mu}\rangle  &=&   \langle T^{\prime\,\,\eta}_{\eta}\rangle + \langle T^{\prime\,\,\chi}_{\chi}\rangle + \langle 
T^{\prime\,\,\theta}_{\theta}\rangle + \langle T^{\prime\,\,\phi}_{\phi}\rangle\nn\\
&=& \frac{{\rm cosech}^8\eta}{240\pi^2}. 
\eea
This is same as (\ref{anom1}).
\bigskip

\noindent{\bf Conclusion:} ~~To conclude, in this short note, we have studied several aspects of the massive and massless scalar fields in 
conformally flat space-times, in particular, 
the FRW geometries. We extended the previous work \cite{Lochan:2018pzs} to include spatially non-flat cases in arbitrary dimensions and with 
non-minimal couplings. We subsequently discussed the restrictions imposed by the energy conditions on the energy-momentum tensors. Further, 
various 
illustrative examples were presented and,  in a radiation dominated FRW universe with a hyperbolic spatial section,
the Feynman propagator of a massless scalar was computed. Finally, we studied relationship between trace anomaly and 
renormalized stress tensors for conformally coupled 
scalars in conformally flat space-times, especially the relationship between the radiation dominated FRW and the Milne space-time.

Several directions for future investigations present themselves. In string theory, besides the massless dilaton and moduli scalars,
various higher spin fields appear. It would be 
interesting to use this technique for these higher spin fields as well. It would also be of interest to apply this technique to areas of 
cosmological perturbations and particle productions in FRW universe.  It would also be interesting to tackle the non-conformal scalar and massive scalar cases and to obtain the relationship between stress tensors for these cases.

\bigskip

\bigskip

\noindent {\bf Acknowledgements:} SMi would like to thank Deepali Mishra and Shuvayu Roy for useful discussions. We thank one of the  referees for suggesting us to 
consider the transformation of the renormalized stress tensor in our context.


\begin{thebibliography}{99}
   
   
\bibitem{Birrell:1982ix} 
  N.~D.~Birrell and P.~C.~W.~Davies,
  ``Quantum Fields in Curved Space,''
  doi:10.1017/CBO9780511622632  
  
\bibitem{parker} 
  L.~Parker  and D.~Toms,
  ``Quantum Field Theory in Curved Space-time,''
  Cambridge University Press, 2009
  https://doi.org/10.1017/CBO9780511813924.  

\bibitem{Lochan:2018pzs}
  K.~Lochan, K.~Rajeev, A.~Vikram and T.~Padmanabhan,
  ``Quantum Correlators in Friedmann Space-times -The omnipresent de Sitter and the invariant vacuum noise,''
  arXiv:1805.08800 [gr-qc].

\bibitem{Iihoshi:2007uz} 
  M.~Iihoshi, S.~V.~Ketov and A.~Morishita,
  ``Conformally flat FRW metrics,''
  Prog.\ Theor.\ Phys.\  {\bf 118}, 475 (2007)
  doi:10.1143/PTP.118.475
  [hep-th/0702139].
  
  

\bibitem{Yamamoto:1994te} 
  K.~Yamamoto, T.~Tanaka and M.~Sasaki,
  ``Particle spectrum created through bubble nucleation and quantum field theory in the Milne Universe,''
  Phys.\ Rev.\ D {\bf 51}, 2968 (1995)
  doi:10.1103/PhysRevD.51.2968
  [gr-qc/9412011].

\bibitem{Sasaki:1994yt} 
  M.~Sasaki, T.~Tanaka and K.~Yamamoto,
  ``Euclidean vacuum mode functions for a scalar field on open de Sitter space,''
  Phys.\ Rev.\ D {\bf 51}, 2979 (1995)
  doi:10.1103/PhysRevD.51.2979
  [gr-qc/9412025].


\bibitem{diSessa:1974ve} 
  A. diSessa,
  "Quantization on hyperboloids and full space-time field expansion,''
  J.\ Math.\ Phys. {\bf 15}, 1892, (1974).

\bibitem{Redmount:1999ps} 
  I.~H.~Redmount,
  ```Natural' vacua in hyperbolic Friedmann-Robertson-Walker space-times,''
  Phys.\ Rev.\ D {\bf 60}, 104004 (1999)
  doi:10.1103/PhysRevD.60.104004
  [gr-qc/9904084].

\bibitem{Bunch:1977sq} 
  T.~S.~Bunch and P.~C.~W.~Davies,
 ``Covariant Point Splitting Regularization for a Scalar Quantum Field in a Robertson-Walker Universe with Spatial 
Curvature,''
  Proc.\ Roy.\ Soc.\ Lond.\ A {\bf 357}, 381 (1977).
  doi:10.1098/rspa.1977.0174

\bibitem{Candelas:1978gf} 
P.~Candelas and J.~S.~Dowker,
  ``Field Theories On Conformally Related Space-times: Some Global Considerations,''
  Phys.\ Rev.\ D {\bf 19}, 2902 (1979).
  doi:10.1103/PhysRevD.19.2902



\end{thebibliography}
\end{document}